\documentclass[prl,aps,twocolumn]{revtex4-1}
\usepackage{amsmath} 
\usepackage{amsthm} 
\usepackage{amssymb}	
\usepackage{graphicx} 
\usepackage[usenames]{color}
\usepackage{array} 
\usepackage{paralist} 
\usepackage{amsfonts}
\usepackage{mathtools}
\usepackage{times}
\usepackage{braket}
\usepackage[colorlinks=true,linkcolor=blue,citecolor=blue,breaklinks]{hyperref}
\normalfont

\makeatletter 
\renewcommand{\thesection}{\normalsize\Roman{section}.}  

\makeatother 
\renewcommand{\vec}[1]{\ensuremath{\mathbf{#1}}} 
 
\let\baraccent=\= 
\renewcommand{\=}[1]{\stackrel{#1}{=}} 

\theoremstyle{definition}

\theoremstyle{remark}

\begin{document}
\newcommand\bbone{\ensuremath{\mathbbm{1}}}
\newcommand{\ul}{\underline}
\newcommand{\vl}{v_{_L}}
\newcommand{\vc}{\mathbf}
\newcommand{\be}{\begin{equation}}
\newcommand{\ee}{\end{equation}}
\newcommand{\bk}{{{\bf{k}}}}
\newcommand{\bK}{{{\bf{K}}}}
\newcommand{\cE}{{{\cal E}}}
\newcommand{\bQ}{{{\bf{Q}}}}
\newcommand{\br}{{{\bf{r}}}}
\newcommand{\bg}{{{\bf{g}}}}
\newcommand{\bG}{{{\bf{G}}}}
\newcommand{\hbr}{{\hat{\bf{r}}}}
\newcommand{\bR}{{{\bf{R}}}}
\newcommand{\bq}{{\bf{q}}}
\newcommand{\hx}{{\hat{x}}}
\newcommand{\hy}{{\hat{y}}}
\newcommand{\hd}{{\hat{\delta}}}
\newcommand{\bea}{\begin{eqnarray}}
\newcommand{\eea}{\end{eqnarray}}
\newcommand{\beal}{\begin{align}}
\newcommand{\eeal}{\end{align}}
\newcommand{\ra}{\rangle}
\newcommand{\la}{\langle}
\renewcommand{\tt}{{\tilde{t}}}
\newcommand{\upa}{\uparrow}
\newcommand{\dna}{\downarrow}
\newcommand{\bS}{{\bf S}}
\newcommand{\vS}{\vec{S}}
\newcommand{\dg}{{\dagger}}
\newcommand{\pdg}{{\phantom\dagger}}
\newcommand{\tphi}{{\tilde\phi}}
\newcommand{\cf}{{\cal F}}
\newcommand{\ca}{{\cal A}}
\renewcommand{\ni}{\noindent}
\newcommand{\ct}{{\cal T}}
\newcommand{\zp}[1]{ { \color{red} \footnotesize ZP\; \textsf{\textsl{#1}} } }

\title{Haldane-Hubbard Mott Insulator: From Tetrahedral Spin Crystal to Chiral Spin Liquid}
\author{Ciar\'{a}n Hickey,$^{1}$ Lukasz Cincio,$^{2}$ Zlatko Papi\'{c},$^{3}$ and Arun Paramekanti$^{1,4}$}
\affiliation{$^1$Department of Physics, University of Toronto, Toronto, Ontario M5S 1A7, Canada}
\affiliation{$^2$Perimeter Institute for Theoretical Physics, Waterloo, Ontario N2L 2Y5, Canada}
\affiliation{$^3$School of Physics and Astronomy, University of Leeds, Leeds, LS2 9JT, United Kingdom}
\affiliation{$^4$Canadian Institute for Advanced Research, Toronto, Ontario M5G 1Z8, Canada}
\begin{abstract}
Motivated by cold atom experiments on Chern insulators, we study
the honeycomb lattice Haldane-Hubbard Mott insulator of spin-$1/2$ fermions using exact diagonalization and density matrix
renormalization group methods. We show that this model exhibits
various chiral magnetic orders including a wide regime of triple-Q tetrahedral order. Incorporating third-neighbor hopping
frustrates and  ultimately melts this tetrahedral spin crystal.
From analyzing the low energy spectrum, many-body Chern numbers, entanglement spectra, and modular
matrices, we identify the molten state as a chiral spin liquid (CSL) with gapped semion excitations.
We formulate and study the Chern-Simons-Higgs field theory of the exotic CSL-to-tetrahedral spin crystallization transition.
\end{abstract}
\maketitle

Electronic bands in crystals can display nontrivial topology, as exemplified by the recent discoveries of
topological insulators \cite{Konig02112007, Hsieh2008}, Weyl semimetals \cite{weyl1,weyl2,weyl3}, and quantum anomalous Hall insulators 
(QAHIs) \cite{Chang2015, Bestwick2015}.
Interactions can dramatically modify this single-particle
physics, for instance by rendering indistinguishable certain topologically distinct free-fermion phases \cite{Fidkowski10,Fidkowski11}.
An alternative outcome is the emergence 
of topological order \cite{Wen95}, manifested by nontrivial 
ground state degeneracies depending on the lattice topology, as discovered in
numerical studies of partially filled Chern bands which realize
lattice fractional quantum Hall liquids \cite{fci_rev1,fci_rev2}. Interactions may also 
lead to charge localization,
while the spin degrees of freedom display topological order. 
Finding even quasi-realistic models of such topological Mott insulators (TMIs) \cite{Pesin2010,Bhattacharjee12,Kargarian2012,Maciejko2014}
is a crucial step towards identifying experimental candidates and understanding
exotic quantum phase transitions out of TMIs.

In this Letter, we study interaction effects in the Haldane model \cite{Haldane1988}, a paradigmatic model of a QAHI 
on the two-dimensional (2D) honeycomb lattice.
The Haldane model supports two bands with Chern numbers $C\!=\! \pm 1$; it has been
realized in recent cold atom experiments \cite{Jotzu2014,Bloch15}.
We study the effect of strong
Hubbard repulsion on spin-$1/2$ (i.e., two-component) fermions in the Haldane model, at a filling 
of one fermion per site, obtaining the following key results.
(i) We establish that 
the effective spin model for the Haldane-Mott insulator exhibits a variety of chiral magnetic orders including a wide
regime of tetrahedral order with large scalar spin chirality. Our results are obtained using exact diagonalization
(ED) on cluster of up to $N\!=\!32$ spins. 
(ii) Incorporating third-neighbor hopping is shown to
frustrate and ultimately melt the tetrahedral order. Our ED results in the liquid phase find a gapped, approximately two-fold degenerate ground 
state, with total many-body Chern number $C\!=\! 1$, suggesting that this state is a chiral spin liquid (CSL):
the $\nu\!=\! 1/2$ bosonic quantum Hall state with gapped semion excitations \cite{AndersonSL, KalmeyerLaughlin, Wen1990}.
We provide conclusive evidence for this using
state-of-the-art density matrix renormalization group (DMRG) \cite{White, mcCulloch08} computations on infinitely long cylinders with circumference 
up to $8$ lattice unit cells, computing entanglement spectra, quantum dimensions of all anyon types, and quasiparticle braiding properties 
via topological $S$ and $T$ matrices.
This frustration-induced
melting of tetrahedral order is a completely {\it distinct} mechanism to realize CSLs compared with previous
studies, and allows us, for the first time, to
identify the tetrahedral state as a `parent' state for the CSL.
(iii) Our ED results suggest a continuous phase transition between the tetrahedral state and the CSL. We formulate
a Chern-Simons-Higgs field
theory to describe this exotic spin crystallization transition out of the topologically ordered CSL.

The study of CSLs was rejuvenated by the construction of exact parent Hamiltonians
\cite{SchroeterCSL2007,ThomaleCSL2009}, and recent works have found evidence for CSLs on the kagome 
\cite{YHe2014,Bauer2014, Gong2014, FradkinCSL2014,Gong2015, Wietek2015,DNShengKagome2015,ShengVMC2015,BieriVWFKag2015,FradkinCSL2015} and
square lattices \cite{Nielsen2013,Poilblanc2015,XJLiu2016}, and in certain $SU(N)$ Mott insulators \cite{Hermele2009} and coupled 
wire models \cite{Meng2015,EranPRB2015}. Our work provides the first example of a CSL on the honeycomb lattice in a realistic model starting from
fermions with on-site interactions. This is nontrivial since a symmetric 
spin-gapped phase on lattices with even number of spin-$1/2$ per unit cell
is not guaranteed to have topological order \cite{OshikawaLSM2000,HastingsLSM2005}.
Our work goes well beyond previous work on this model
\cite{He2011_1,He2011_2,Maciejko13,Hickey2015,Zhai2015}, and studies of Gutzwiller projected Chern-insulator
wavefunctions \cite{Zhang2011, Zhang2012} which did not consider microscopic models that support such ground states.
The tetrahedral state we find here also occurs in certain triangular lattice Hubbard and Kondo models \cite{Martin08,Jiang15},
suggesting that such frustration-induced CSLs may appear in a wider class of models and materials.

\begin{figure}[t]
\includegraphics[width=\linewidth]{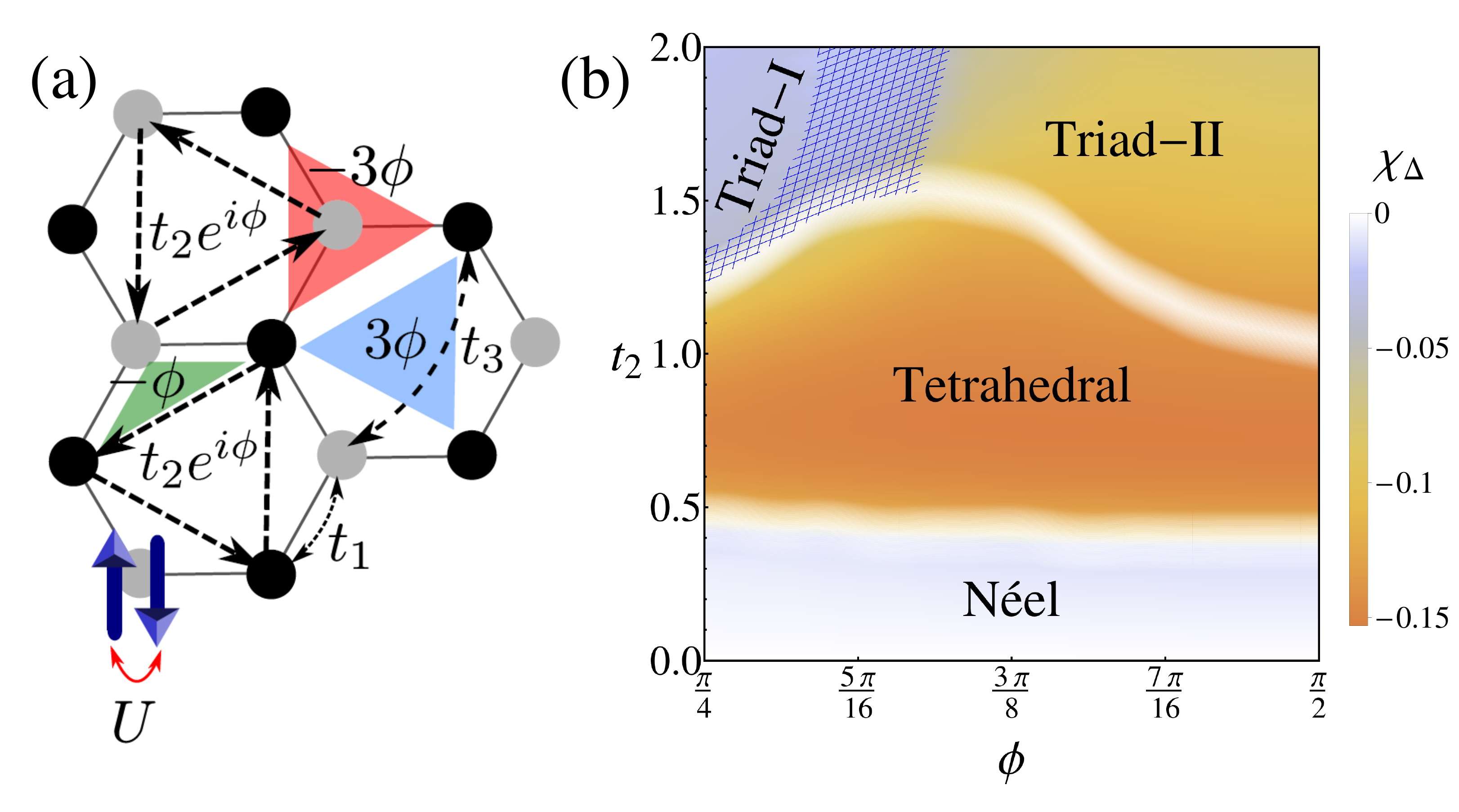}
\caption{(Color online) (a) Haldane-Hubbard model showing short distance hopping amplitudes, plaquette fluxes, and
Hubbard repulsion $U$.
(b) Phase diagram of $H_{\rm spin}$ for $t_3\!=\! 0,U\!=\!10$ from ED on clusters with $N\!=\! 24$ spins; color 
indicates the chirality $\la \hat{\chi}_\triangle\ra$ on small triangles. White lines indicate phase boundaries, 
broadened to account for finite-size effects. In the hatched (blue) region we cannot sharply identify the phase in ED as Triad-I or II.}
\label{fig:ed1}
\end{figure}

\ni {\bf Model.} The Haldane-Hubbard model for spin-$1/2$ fermions shown in Fig.~\ref{fig:ed1}(a) is defined by the Hamiltonian
\begin{align}
\notag H_{\rm HH} \!\! =\!\!& -t_1\! \!\! \sum_{\la i j \ra \sigma} (c^\dg_{i\sigma} c^\pdg_{j\sigma} \!+\! h.c.) 
\!-\!  t_2 \!\!\!\! \sum_{\la\la i j \ra\ra \sigma} \!\! (e^{i\nu_{ij}\phi} c^\dg_{i\sigma} c^\pdg _{j\sigma} \!+\! h.c.)  \\
&+ U \sum_{i} n_{i\uparrow}n_{i\downarrow},   \label{eqn:HaldaneHubbard}
\end{align}
where $\la.\ra$ and $\la\la.\ra\ra$ denote, respectively, first and second nearest neighbors, $\nu_{ij}\!=\! \pm 1$
produces a flux pattern with a net zero flux per unit cell, and $U$ is the Hubbard repulsion. 
For $U\!=\! 0$, this model supports Chern bands for $t_2,\phi \!\neq \! 0$. At half-filling, this leads to a 
QAHI with $\sigma_{xy}\!=\! \pm e^2/h$ per spin
for small $|t_2|$. At large $|t_2|$ and $\phi\neq\pi/2$, the Chern bands strongly disperse, leading to a metal with 
$\sigma_{xy} \neq 0$ but non-quantized \cite{Hickey2015}.

For $U \! \gg \! |t_{1,2}|$, degenerate perturbation theory in the Mott insulator \cite{Yoshioka1988} with one fermion
per site
leads to the spin model
\begin{align}
\!\! \notag H_{\rm spin} \!&= \frac{4 t_1^2}{U} \!\! \sum_{\la ij \ra} \vec S_i \cdot \vec S_j
 + \frac{4t_2^2}{U} \!\! \sum_{\la\la ij \ra\ra} \!\! \vec S_i \cdot \vec S_j  \\
\!\!\!&+\frac{24t_1^{2}t_2}{U^2} \!\!\!\!\!\! \sum_{{\rm small}-\triangle} \!\!\!   \hat{\chi}_\triangle \sin\Phi_\triangle 
+ \frac{24t_2^{3}}{U^2} \!\!\! \sum_{{\rm big}-\triangle} \! \hat{\chi}_\triangle \sin \Phi_\triangle,
\label{Hspinfermion}
\end{align}
where $\hat{\chi}_\triangle \equiv \vec S_i \cdot (\vec S_j \times \vec S_k)$ is the scalar spin chirality operator.
The sites $\{ijk\}$ in $\hat{\chi}_\triangle$ are labelled going anticlockwise around the small or big triangles of the honeycomb 
lattice. As shown in Fig.~\ref{fig:ed1}(a), the fluxes in $H_{\rm spin}$ are $\Phi_\triangle=-\phi$ on small (green) triangles,
and $\Phi_\triangle=-3 \phi$ $(+3\phi)$ on large triangles which do (do not) enclose
a lattice site. Classical
magnetic ground states of this model, valid for $S\!=\! \infty$, have been studied in \cite{Hickey2015};  here, 
we resort to a numerical study for $S\!=\!1/2$, retaining strong quantum fluctuations.

\noindent {\bf ED phase diagram.} For $\phi=0$, $H_{\rm spin}$ reduces to the $J_1$-$J_2$ honeycomb lattice
Heisenberg model, with $J_{1,2}=4 t^2_{1,2}/U$.
Previous work indicates that $J_2 \gtrsim 0.2 J_1$ kills N\'eel order, 
leading to incommensurate spirals \cite{Mulder2010} for $S\!=\! \infty$, and competing 
valence bond crystals for $S\!=\!1/2$ \cite{Fouet2001,Albuquerque2011,Mosadeq2011}. Here, we study the
unexplored regime $\phi \neq 0$, using Lanczos ED on clusters up to $N=32$ spins, varying $t_2$ and $\phi$ for fixed $U/t_1=10$ which puts
us in the Mott insulator \cite{Hickey2015}. We focus on flux values $\pi/4 \leq \phi \leq \pi/2$, 
which reveals commensurate phases with large scalar spin chirality;
restricting ourselves to this window of flux avoids  incommensurate spiral orders \cite{Mulder2010,Hickey2015} expected at small $\phi$,
which have strong finite-size effects in ED.
Below, we work in units where $t_1\!=\! 1$.

As shown in Fig.\ref{fig:ed1}(b), we find that the phase diagram contains four magnetically ordered phases --- N\'eel, tetrahedral and triad-I/II orders --- which
are also observed in the classical phase diagram \cite{Hickey2015}. 
(i) The N\'eel order on the honeycomb lattice is translationally invariant, with ferromagnetic order on each sublattice and a single structure factor peak at the $\Gamma$ point
of the hexagonal Brillouin zone.
(ii) The tetrahedral order has an $8$-site magnetic unit cell, with spins pointing toward the four corners of a tetrahedron and structure factor peaks at the three $M$ points. It 
is a so-called ``regular magnetic order", respecting all lattice symmetries modulo global spin rotations.
(iii)/(iv) Triad-I/II both have $6$-site magnetic unit cells, with 
three spins on each sublattice forming a cone and structure factor peaks at the $K$ and $K^\prime$ points. They can be thought of as umbrella states 
on each triangular sublattice, with their common axis being parallel in the triad-I case and anti-parallel in the triad-II. This yields a net ferromagnetic moment 
in triad-I and a net staggered moment in
triad-II.

We identify these magnetic orders within ED, on clusters with up to $N\!=\! 32$ spins, through a careful analysis of the low energy spectrum, extracting quantum numbers of the
quasi-degenerate joint states, i.e., the `Anderson tower', in each total spin sector, whose energies collapse onto the ground state as $1/N$
leading to spontaneous symmetry breaking in the thermodynamic limit \cite{LhuillierQDJS1992,LhuillierSpectra1994} (see Supplemental Material \cite{SuppMat}). The 
phase boundaries in Fig.\ref{fig:ed1}(b) are determined \cite{SuppMat} by
dips in the ground state fidelity $\braket{\Psi_0(g) | \Psi_0(g \! +\! \delta g)}$ which signal quantum phase transitions \cite{Fidelity2006}, where $g$ is a tuning parameter
(here, $t_2$ or $\phi$). We substantiate this by studying changes in 
the finite-size singlet ($E_s$) and triplet ($E_t$) gaps, $\la \hat{\chi}_\triangle \ra$, and reorganization of the low energy spectrum.
Our results are in contrast to slave-rotor mean field theory of the 
Haldane Mott insulator \cite{He2011_1,He2011_2}, in which the ground state is a CSL which simply inherits
the band topology of the underlying QAHI.

\begin{figure}[thb]
\includegraphics[width=\linewidth]{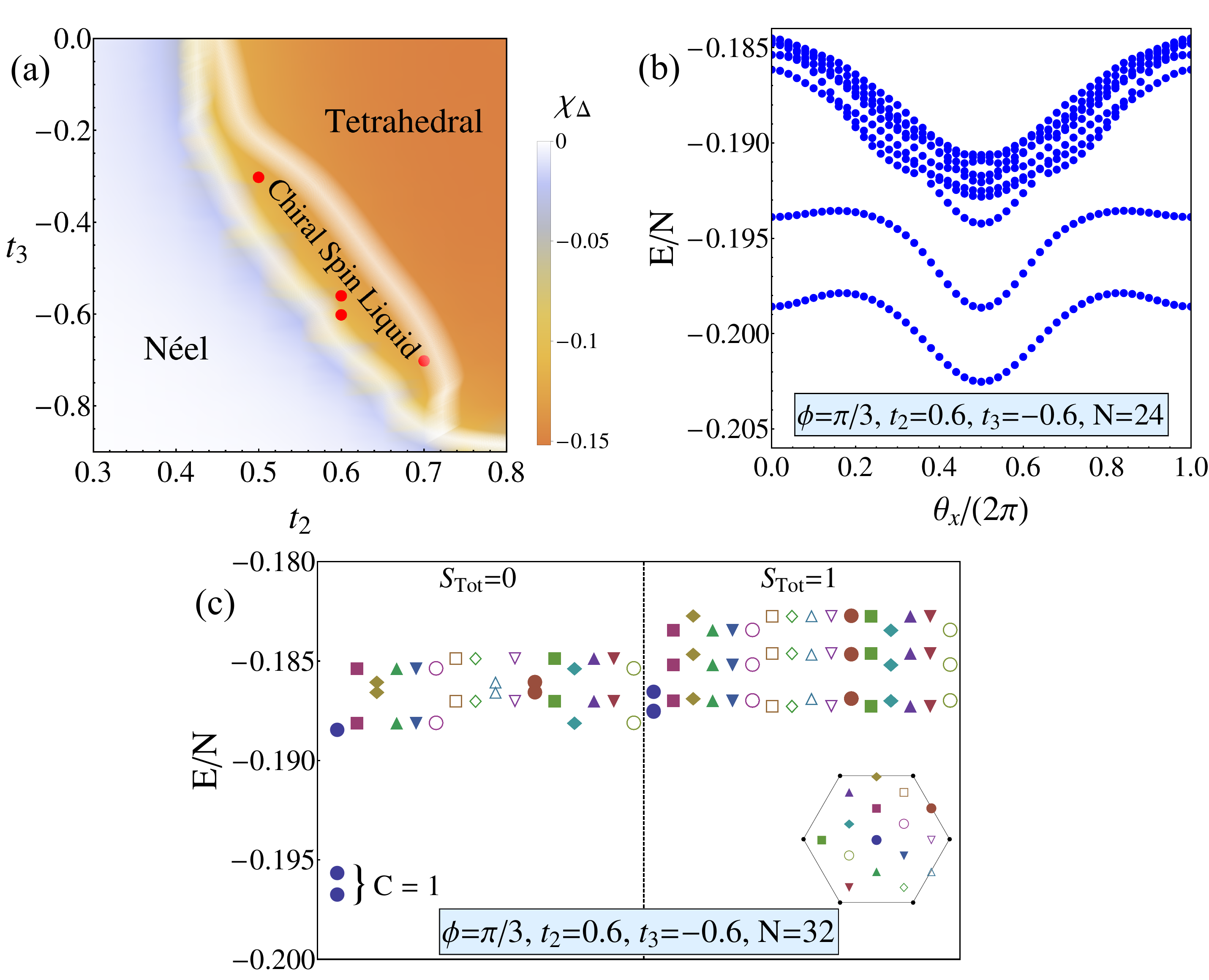}
\caption{(Color online) (a) Phase diagram of $H_{\rm spin}$
at $\phi\!=\! \pi/3$ and $U=10$, keeping the additional $J_3$ term induced by $t_3 \neq 0$. Background
shows ground state chirality $\la \hat{\chi}_\triangle \ra$ on
small-$\triangle$. Using ED
and DMRG (at indicated points), we find a window of CSL with topological order. (b) Topological robustness of the CSL ground states upon threading flux through one hole of the torus. Energy spectrum as a function of boundary phase $\theta_x$ is shown for $N\!=\! 24$ sites, $t_2\!=\! 0.6$, and $t_3\!=\! -0.6$. 
(c) Energy spectrum for $N\!=\! 32$ cluster, with states labelled by total spin $S_{\rm tot}$ and Brillouin zone momenta shown in the inset. 
We find approximate two-fold ground state degeneracy with total Chern number $C_1+C_2\!=\! 1$.}
\label{fig:ed2}
\end{figure}

\noindent {\bf Melting tetrahedral order.} The tetrahedral state is a ``regular magnetic state" \cite{Misguich11} which respects all lattice symmetries in its $SU(2)$-invariant correlations.
Given its large scalar spin chirality, it is tempting to speculate that quantum disordering this state might lead to
a CSL. We thus modify the
Haldane model in order to frustrate the tetrahedral order.
We notice that the tetrahedral state has spins on opposite vertices of the honeycomb hexagon aligned ferromagnetically.
Thus incorporating third-neighbor hopping $t_3$ will lead to an additional exchange interactions in $H_{\rm spin}$, i.e.,
the Heisenberg exchange $J_3\!=\! 4 t^2_3/U \!>\!  0$ which will inevitably frustrate tetrahedral order, as well as additional chiral interactions. 
Below, we present extensive results retaining only $J_3>0$ since keeping all chiral terms induced by $t_3$ significantly
increases the computational complexity; we have explicitly checked that these additional terms induce very small quantitative
differences in the ED spectra, and only slightly shift the phase boundaries in the phase diagram (see Supplemental Material \cite{SuppMat}).

One key signature of a CSL is a nonzero spin gap and two-fold ground state degeneracy on the torus. We thus look for regimes 
where the lowest excited state is a spin-singlet whose energy
gap becomes smaller with system size, while the triplet gap remains nonzero. Fig.~\ref{fig:ed2}(a) shows the ED phase diagram as we 
vary $(t_2,t_3)$, where we find a candidate CSL regime. Here, we have fixed $\phi\!=\! \pi/3$, at which the coefficient of 
$\hat{\chi}_\triangle$ on the large-$\triangle$ vanishes, enormously simplifying the numerics.

Fig.~\ref{fig:ed2}(c) shows a representative ED spectrum on an $N\!=\! 32$ torus at $(t_2,\! t_3) \!=\! (0.6,\! -0.6)$. We find an
approximate two-fold ground state degeneracy, both states being spin singlets
with crystal momentum $\bk\!=\! (0,0)$ as expected for a honeycomb lattice CSL,
and a spin gap $E_t \! \approx \! 0.3$. Threading flux through one hole of the torus (see Fig.~\ref{fig:ed2}(b)), 
we find the two-fold ground state manifold does not with mix with
higher excited states, demonstrating that the ground state degeneracy is of topological origin. 
We have computed the many-body Chern numbers
$C_i \!=\! - \frac{1}{\pi} \int\!\! d {\theta_1} d\theta_2 {\rm Im}\Braket{\partial_{\theta_1}\Psi_i | \partial_{\theta_2}\Psi_i }$
using twisted boundary conditions on the two ground states $|\Psi_{i=1,2}\ra$,
since two ground states have the same momentum and thus do not cross. However, only the total Chern number
of this degenerate manifold is meaningful in the thermodynamic limit;
we find $C_1\!+\!C_2\!=\! 1$. 
These results provide strong evidence that $t_3$ melts tetrahedral order, leading to a  $\nu\!=\!1/2$ bosonic 
Laughlin liquid. Our ED results delineate a regime at $\phi\!=\! \pi/3$, see Fig.~\ref{fig:ed2}(a), 
which we identify as a CSL candidate.

\begin{figure}[!t]  
\begin{center}
\includegraphics[width=\columnwidth]{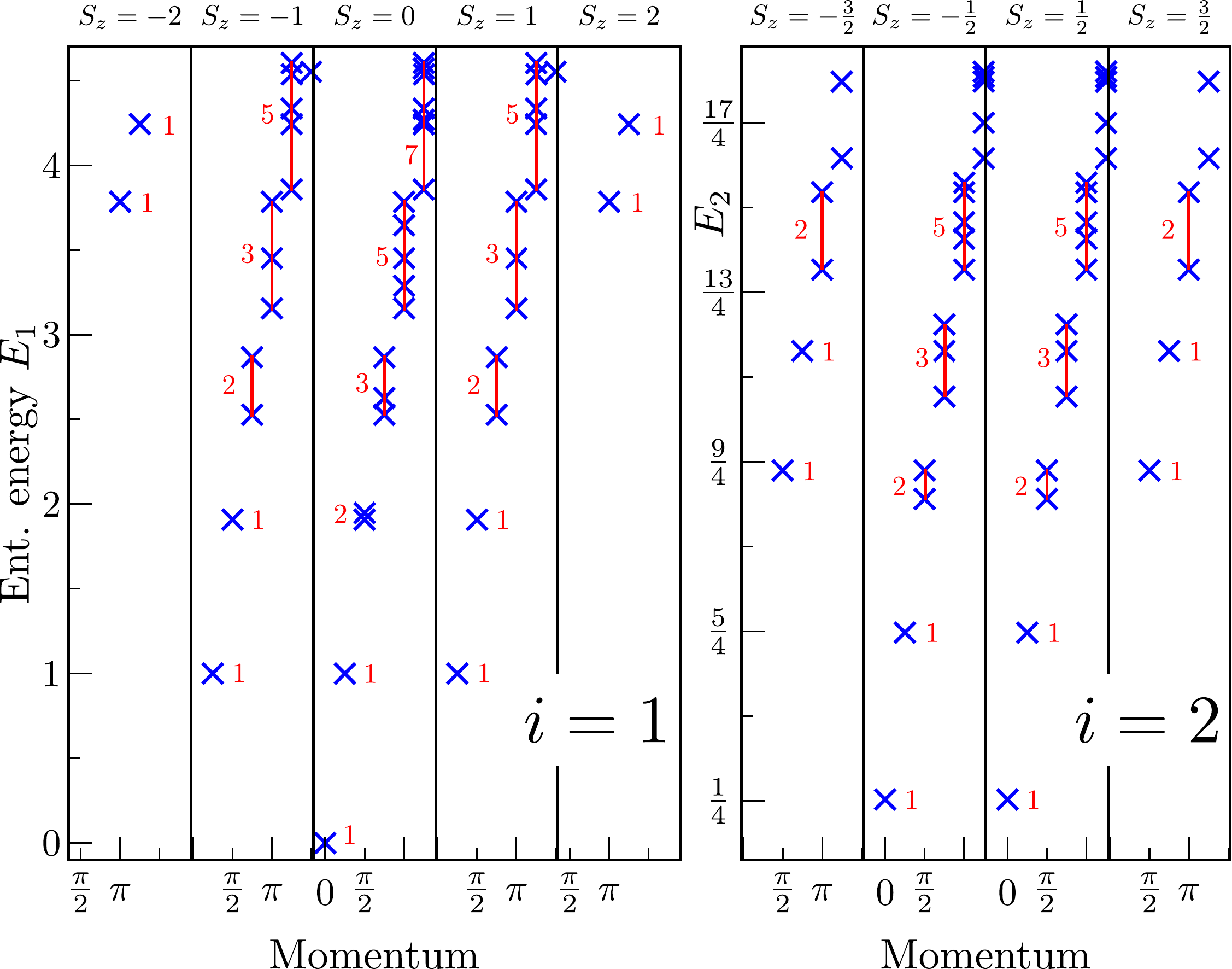}
\end{center}
\caption{(Color online) Entanglement spectrum (rescaled and shifted) of the reduced density matrix $\rho_i$ for half an infinite cylinder (with circumference $L=8$
unit cells) computed for the ground states $| \Psi_1^\mathrm{cyl} \rangle$ (left panel) and $| \Psi_2^\mathrm{cyl} \rangle$ (right panel) of the effective spin model at $(t_2,t_3,\phi) = (0.6,-0.6,\pi/3)$. Vertical axes show entanglement energies defined as $E_{i,\alpha} = -\log \lambda_{i,\alpha}$, where $\lambda_{i,\alpha}$ are the eigenvalues of $\rho_i$. The transverse momenta of the corresponding eigenvectors of $\rho_i$ are shown on horizontal axes, separately for every tower labeled by $S_z$ quantum number. The number of close lying states with the same momenta in a given $S_z$ sector is shown in red.}
	\label{fig:EntSpec}
\end{figure}

\noindent {\bf DMRG results.} To further confirm the existence of CSL, we investigate the model $H_{\rm spin}$ with additional terms generated by non-zero $t_3$,
using DMRG \cite{mcCulloch08}, on a cylinder of infinite length with circumference up to $L\!=\!8$ unit cells.
The characterization of a topologically ordered phase is achieved by: (i) identifying the conformal field theory (CFT) that describes gapless edge excitations via the ``entanglement spectrum" \cite{li08}, and 
(ii) computing topological $S$ and $T$ matrices that contain information about bulk anyon excitations \cite{Wen1990,Zhang2012, Rowell09, Cincio2013, Zaletel2013, Zhu2013}.
Simulations were performed for $\phi\!=\! \pi/3$, and four different values of $(t_2,t_3)$ marked by red dots on the phase diagram in Fig.\ref{fig:ed2}(a), keeping only the
additional $J_3$ exchange term. We present detailed results below for one point $(t_2,t_3)\! =\! (0.6,-0.6)$; we obtain similar results at the other three points. 
We also performed simulations on smaller width cylinders (upto $L=6$) keeping $J_3$ {\it and} all additional chiral terms from having $t_3 \!\neq \!0$ 
in $H_{\rm HH}$, obtaining similar results.

Randomly initialized DMRG finds two ground states, $| \Psi_{i=1,2}^\mathrm{cyl} \rangle$, with well-defined anyon flux threading inside the cylinder \cite{Cincio2013}. Fig. \ref{fig:EntSpec} shows the entanglement spectrum $E_i$ of the reduced density matrix for half an infinite cylinder computed for both ground states. Studying these spectra, we can extract universal information about possible gapless boundary excitations, as if the system had an actual, physical edge \cite{li08, chandran11, dubail12, swingle12, qi12}. The spectra $E_i$ are seen to be consistent with corresponding sectors of the chiral $SU(2)_1$ Wess-Zumino-Witten CFT \cite{Wen1990CLL}. $E_1$ is associated with the identity primary operator and its Kac-Moody descendants.
The computed degeneracy pattern in every tower (labeled by $S_z$) is seen to follow the expected partition numbers (1--1--2--3--5--7--...) \cite{difrancesco97}. $E_2$ corresponds to the chiral boson vertex operator and its descendants.

The ground states $| \Psi_{i=1,2}^\mathrm{cyl} \rangle$ on an infinite cylinder $\infty \times L$ may be used to mimic grounds states on a $L \times L$ torus $| \Psi_{i=1,2}^\mathrm{tor} \rangle$ by means of cutting and reconnecting matrix-product states of $| \Psi_i^\mathrm{cyl} \rangle$ \cite{Cincio2013, Zaletel2013}. Every such ground state $| \Psi_i^\mathrm{tor} \rangle$ has a well-defined anyon flux threading inside the torus. The topological $S$ and $T$ matrices of the emergent anyons can be extracted \cite{zhang12} from the overlaps $\langle \Psi_i^\mathrm{tor}| R_{\pi/3} | \Psi_j^\mathrm{tor} \rangle$, where $R_{\pi/3}$ denotes clockwise $\pi/3$ rotation of a $L \times L$ torus. For $L=6$, we find
\begin{eqnarray}
S &=& \frac{1}{\sqrt{2}} \left( \begin{matrix} 
0.99 & 0.97 \\
0.96 & -0.97 \cdot e^{i\pi \cdot 0.01}
\end{matrix} \right), \\ 
T &=& e^{i \frac{2\pi}{24} \cdot 0.96} \left( \begin{matrix}
1 & 0 \\
0 & -i \cdot e^{i\pi\cdot 0.01}
\end{matrix} \right),
\end{eqnarray}
in excellent agreement with the exact $S$ and $T$ matrices of a chiral semion anyon model, $\frac{1}{\sqrt{2}} \left( \begin{matrix} 1 & 1 \\ 1 & -1\end{matrix}\right)$ and $e^{i\frac{2\pi}{24}} \left(\begin{matrix} 1 & 0 \\ 0 & -i \end{matrix}\right)$. The combined DMRG results thus provide an unambiguous identification of the phase as a CSL.

\ni {\bf Spin crystallization transition.} Our ED results show that the chirality and ground state 
fidelity vary smoothly going from the tetrahedral state into the CSL. This suggests that the two phases
might be separated by an exotic critical point since the tetrahedral state is topologically trivial but
breaks $SU(2)$ spin symmetry while the CSL has topological order and no broken symmetries.
A powerful route to accessing such exotic transitions is via fractionalizing the spins \cite{Senthil04}. We formulate our theory in terms of 
spin-$1/2$ bosonic spinons minimally coupled to an Abelian level $k\!=\!2$ Chern-Simons (CS) gauge field. In the CSL,
integrating out gapped
spinons results in a CS topological field theory. The lowest energy excitations are gapped spinons,
which carry unit gauge charge and bind $\pi$-flux, converting them into semions. On the tetrahedral side,
spinon condensation produces magnetic order, destroying topological
order via the Higgs mechanism.

To construct the field theory for the matter
sector, we imagine bosonic spinons with spins polarized along the local Zeeman axes of the underlying tetrahedral order.
Adiabatic spinon transport around closed loops on the honeycomb lattice then produces nontrivial Berry phases; we
find $\pi$-flux around hexagonal loops and $\pi/2$-flux around
triangular plaquettes. Even if long wavelength quantum fluctuations disorder the tetrahedral state, so these Zeeman fields
average to zero, we expect 
the local spin chirality and hence the local fluxes to persist. Diagonalizing this spinon Hofstadter Hamiltonian on the honeycomb lattice, 
we find $4$ equivalent dispersion minima located, for our gauge choice, at $\bQ_0 \equiv \Gamma$ and $\bQ_i \equiv M_i$ 
($i\!=\!1,2,3$; the three $M$ points of the BZ).
We thus study the action $S \!=\! \int d^2xd\tau ({\cal L}_{\rm CS,\phi} + {\cal L}_{\rm int}  )$, where
\bea
{\cal L}_{\rm CS,\phi} \! &=& \!\! \frac{1}{2\pi} \epsilon^{\mu\nu\lambda} a_\mu \partial_\nu a_\lambda + |(\partial_\mu \!-\! i a_\mu) \phi^\pdg_{i\alpha}|^2
+ r |\phi^\pdg_{i\alpha}|^2
\eea
describes bosonic spinons minimally coupled to the CS gauge field, while ${\cal L}_{\rm int}={\cal L}^{(1)}_{\rm int}+{\cal L}^{(2)}_{\rm int}$ captures spinon interactions,
\bea
{\cal L}^{(1)}_{\rm int} \! &=& \! u_1 (\sum_i \rho_i)^2
\!+\! u_2 \! \sum_{i\neq j}  \rho_i \rho_j
\!+\! u_3 \! \sum_{i\neq j}  \roarrow {\cal S}_i \cdot \roarrow {\cal S}_j \nonumber \\
&+& u_4 \!\! \sum_{[ijk\ell]}
\phi^*_{i\alpha} \phi^*_{j\beta}\phi^\pdg_{k\alpha}\phi^\pdg_{\ell\beta}
\!+\! u_5\! \sum_{i\neq j}  \phi^*_{i\alpha} \phi^*_{i\beta}\phi^\pdg_{j\alpha}\phi^\pdg_{j\beta}  \nonumber\\
{\cal L}^{(2)}_{\rm int} &=& w_1 (\sum_i \rho_i)^3 + w_2 \sum_{i,j,k} \epsilon^{ijk} \roarrow {\cal S}_i\cdot (\roarrow {\cal S}_j \times \roarrow {\cal S}_k) + \ldots
\eea
Latin indices label the $4$ modes at $\bQ_i$  ($i=0,1,2,3$), the notation $[ijk\ell]$ implies all $4$ modes are different,
and there is an implicit sum on Greek indices which label spin or space-time.
We defined
$\rho_i \equiv \phi^*_{i\alpha} \phi^\pdg_{i\alpha}$ and 
$\roarrow {\cal S}_i \equiv \phi^*_{i\alpha} \roarrow \sigma^\pdg_{\alpha\beta} \phi^\pdg_{i\beta}$.
${\cal L}^{(1)}_{\rm int}$ and ${\cal L}^{(2)}_{\rm int}$ respectively list all quartic interactions and important sixth order terms, consistent with
momentum conservation, global $SU(2)$ symmetry, and local gauge invariance. $u_{1,2}$
are forward-scattering interactions, $u_{3,4}$ are backscattering terms, and $u_{5}$ is an Umklapp process.
$w_2$ encodes broken time-reversal symmetry. At mean field level,
with dominant $u_1,w_1 \! > \! 0$, we find $ r  \!>\! 0$ leads to the CSL, while
tuning $r \!<\! 0$ leads to a confining Higgs phase with $\la \phi_{i\alpha} \ra \! \neq \! 0$.
For $u_2\!<\! 0$, we get simultaneous condensation at all $\bQ_i$. The tetrahedral state emerges via a continuous transition for
subdominant terms $u_4,u_5 \! < \! u_3,w_2$ (see Supplemental Material \cite{SuppMat}).
Our construction of the field theory for the CSL-tetrahedral transition relies on a nontrivial flux pattern for
the spinons, hinting at 
`crystal symmetry fractionalization' \cite{Essin2014} in the CSL.

\noindent{\bf Summary.} Using ED and DMRG, we have shown that the Haldane-Hubbard Mott insulator supports unusual chiral magnetic orders, while 
third-neighbor hopping induces a CSL with topological order. 
We have argued that this CSL descends from a `parent' tetrahedral state and 
constructed a CS-Higgs theory for this exotic spin-crystallization transition. Recent work has shown that the kagome lattice admits only a 
single $SU(2)$ invariant symmetry enriched CSL \cite{WhiteCSF2015,LukaszCSF2015}. However, the
honeycomb lattice may admit multiple CSLs with distinct crystal symmetry fractionalization
patterns. Future research
directions include nailing down the precise nature of this CSL 
\cite{Motrunich15,Zaletel15,FuCSF2015,WhiteCSF2015,LukaszCSF2015}, and 
relating this CSL to Gutzwiller 
projected wavefunctions \cite{Zhang2011, Zhang2012}. Another outstanding issue is fluctuation effects
on the CS-Higgs transition proposed here,
and in related U(1) symmetric bosonic quantum Hall to charge density-wave insulator transitions \cite{Barkeshli2015}.

\noindent{\bf Acknowledgments.} We thank R. Desbuquois, K. Hwang, G. Jotzu, C. Laumann, 
S. Sachdev, A. Thomson, S. Whitsitt, and D.N. Sheng for useful discussions. CH and AP acknowledge support from NSERC of Canada. Computations were performed on the GPC supercomputer at the SciNet HPC Consortium. SciNet is funded by: the Canada Foundation for Innovation under the auspices of Compute Canada; the Government of Ontario; Ontario Research Fund - Research Excellence; and the University of Toronto. This work also made use of the facilities of N8 HPC Centre of Excellence, provided and funded by the N8 consortium and EPSRC (Grant No.EP/K000225/1). The Centre is co-ordinated by the Universities of Leeds and Manchester. L.C. acknowledges support by the John Templeton Foundation. This research was supported in part by Perimeter Institute for Theoretical Physics. Research at Perimeter Institute is supported by the Government of Canada through Industry Canada and by the Province of Ontario through the Ministry of Research and Innovation.

\vfill\eject

\widetext

\appendix
\newpage
\section{Supplemental Material}

\renewcommand{\theequation}{S\arabic{equation}} 
\renewcommand{\thesection}{S\arabic{section}}  
\renewcommand{\thetable}{S\arabic{table}}  
\renewcommand{\thefigure}{S\arabic{figure}}

\section{Exact diagonalization spectra for magnetically ordered states with $t_3=0$}

Exact diagonalization (ED) on system sizes of up to $N=32$ spins is used to construct the phase diagram of the Haldane-Hubbard Mott insulator, with fixed $U=10$ and 
varying $t_2$ and flux $\phi$. The magnetic orders present can be identified by analysing the quantum numbers of the low-lying states in each total spin sector of the ED energy spectrum, the so-called
`quasi-degenerate joint states' (QDJS), or `Anderson tower'. These states collapse onto the ground state as $1/N$
leading to a spontaneous symmetry broken ground state in the thermodynamic limit.

\begin{figure}[thb]
\begin{center}
\includegraphics[scale=0.18]{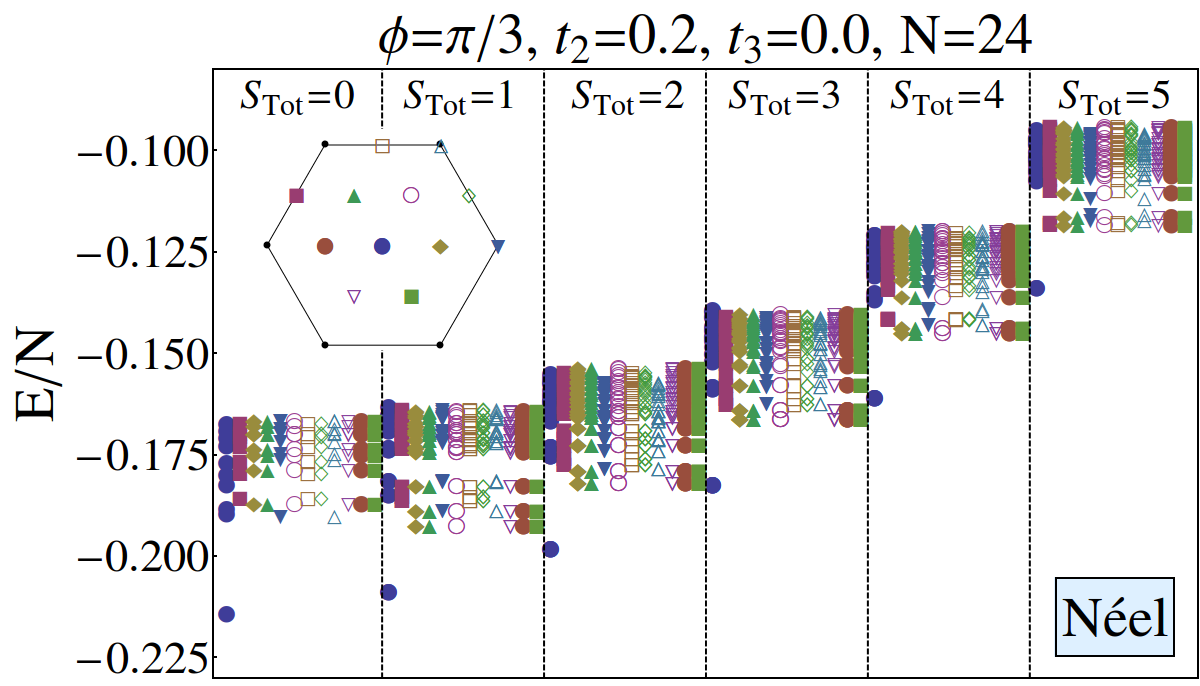}
\includegraphics[scale=0.18]{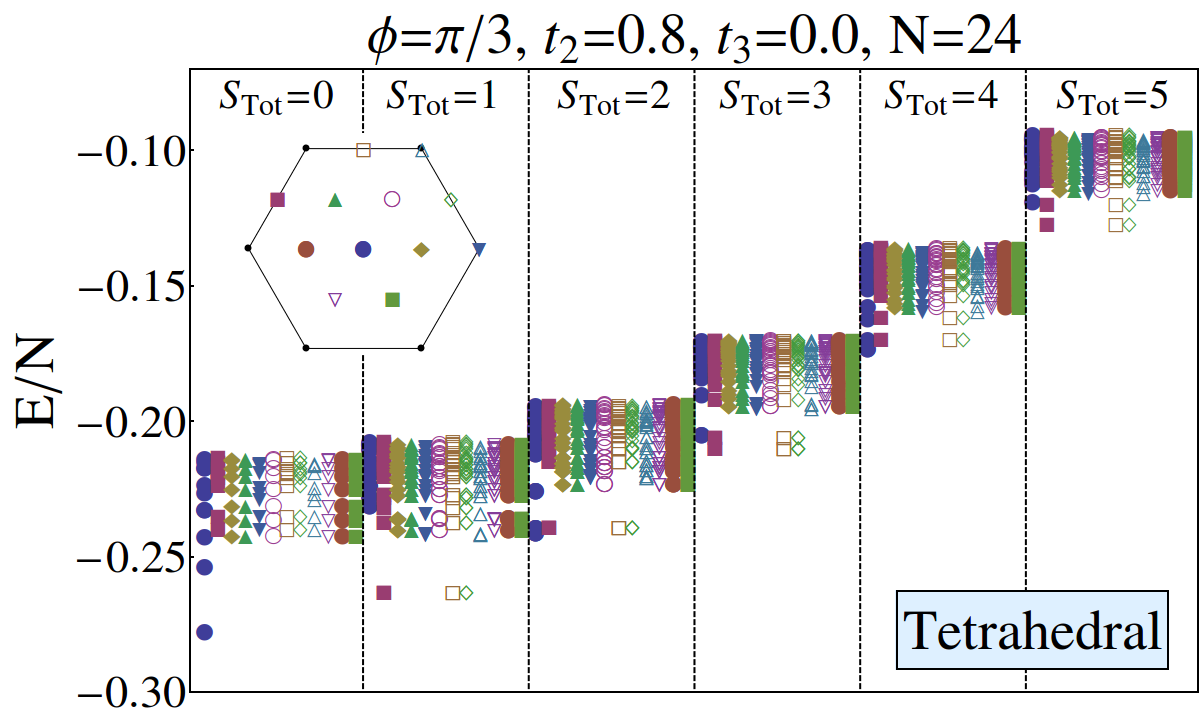}
\includegraphics[scale=0.18]{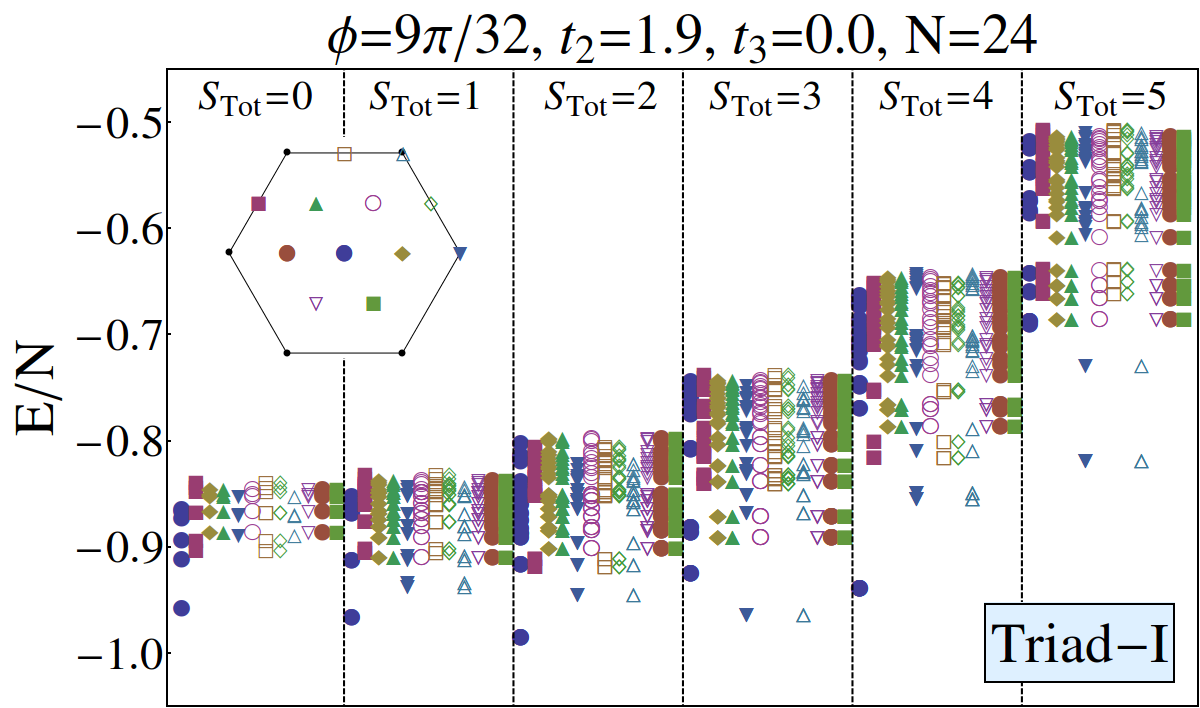}
\includegraphics[scale=0.18]{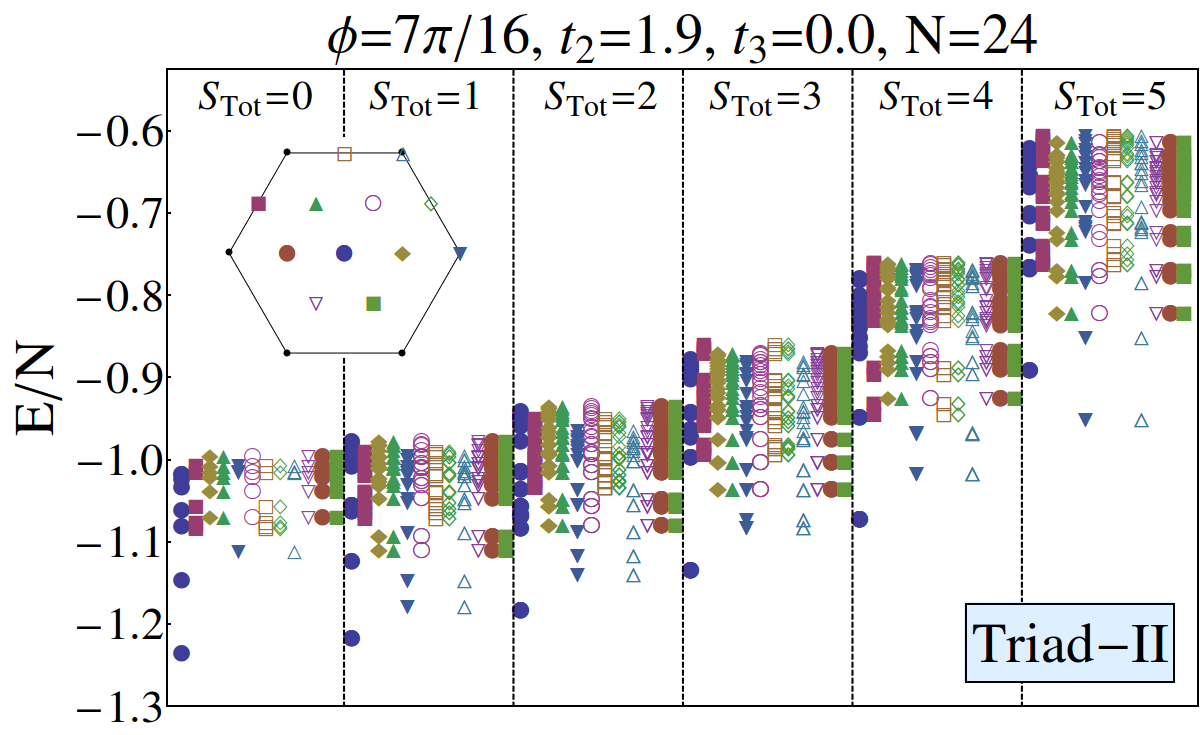}
\caption{Example ED energy spectra for the four magnetically ordered phases described in the main text for a $N=24$ spin cluster.}
\label{fig:ED_Spectra}
\end{center}
\end{figure}

As stated in the main text, we find that the phase diagram contains four magnetically ordered phases - N\'{e}el, tetrahedral and triad-I/II orders. In Fig. \ref{fig:ED_Spectra} we present example spectra for these four phases for a $N=24$ site cluster which has the full point group symmetry of the lattice $C_{6v}$. In this case the QDJS can be characterised by their momenta and irreducible representation (IR) of $C_{6v}$. 
Properties of the phases include:\\
$\bullet$ The N\'eel order is collinear and translationally invariant, with QDJS with momentum at the $\Gamma$ point and energy scaling linearly with $S_{Tot}(S_{Tot}+1)$ as 
expected for quantum rotor excitations. \\
$\bullet$ The tetrahedral order is non-coplanar with QDJS with momentum at the $\Gamma$ and $M$ points and large chirality on small triangles.
$\bullet$ The triad-I order is non-coplanar and has a net ferromagnetic moment (with the ground state lying in a sector with $S_{Tot}\neq0$), with QDJS at the $\Gamma$ point and the $K, K^\prime$ 
points as expected.  \\
$\bullet$ The triad-II order is similar in many respects to the triad-I but with a net anti-ferromagnetic moment and oppositely signed chirality on big triangles.

\section{Ground state fidelity exact diagonlization results}
\begin{figure}[thb]
\includegraphics[scale=0.3]{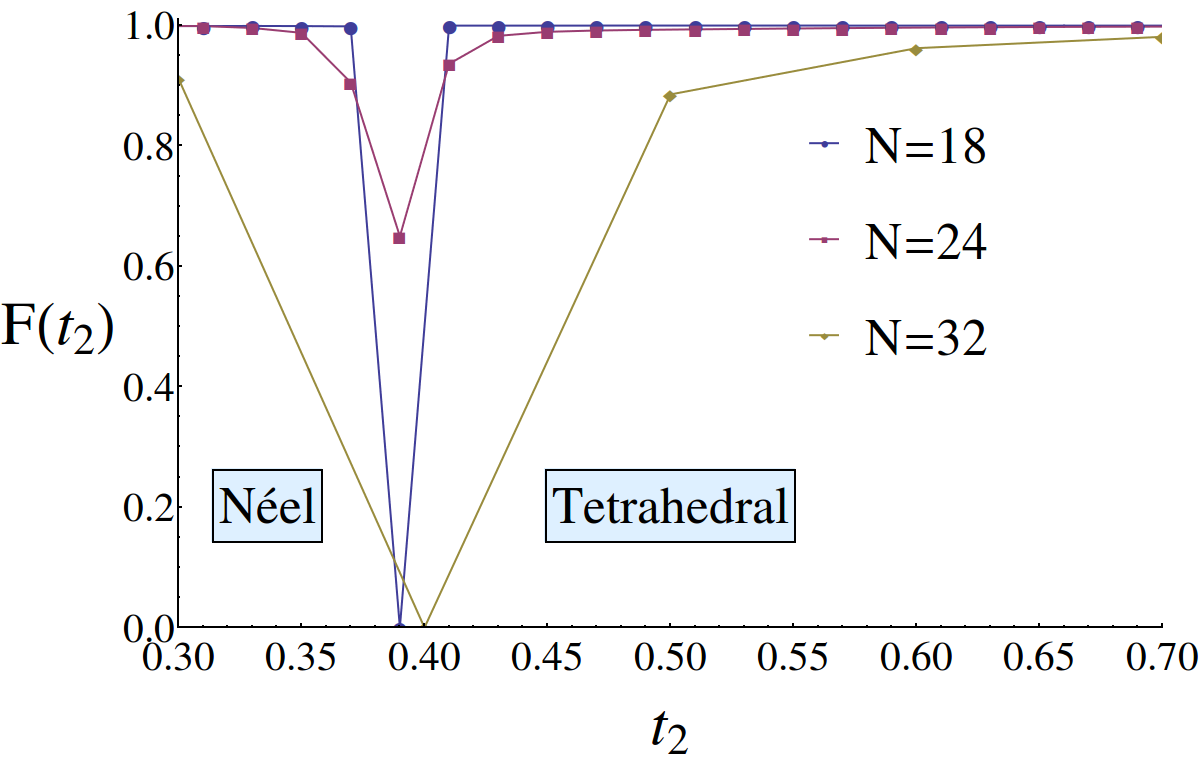}
\caption{Ground state fidelity results, $F(t_2)=\braket{\Psi_0(t_2) | \Psi_0(t_2 \! +\! \delta t_2)}$, for $N=18,24$ and $32$ site torus geometries at $\phi=\pi/2, t_3=0$. }
\label{fig:Fid}
\end{figure}

The phase boundaries were determined by analysing dips in the ground state fidelity, $F(g)=\braket{\Psi_0(g) | \Psi_0(g \! +\! \delta g)}$ with $g$ a tuning parameter, as well as changes in the low energy spectrum, the finite-size singlet ($E_s$) and triplet ($E_t$) gaps and the scalar spin chirality $\la \hat{\chi}_\triangle \ra$ on big and small traingles. In Fig. \ref{fig:Fid} we show the ground state fidelity as a function of $t_2$ for $N=18,24$ and $32$ site torus geometries at $\phi=\pi/2, t_3=0$. The sharp dips mark the transition from the N\'{e}el to the tetrahedral state.

\section{Comparison between ED results for $t_3 \neq 0$ for the full model versus simplified model with only $J_3>0$}

\begin{figure}[thb]
\includegraphics[scale=0.33]{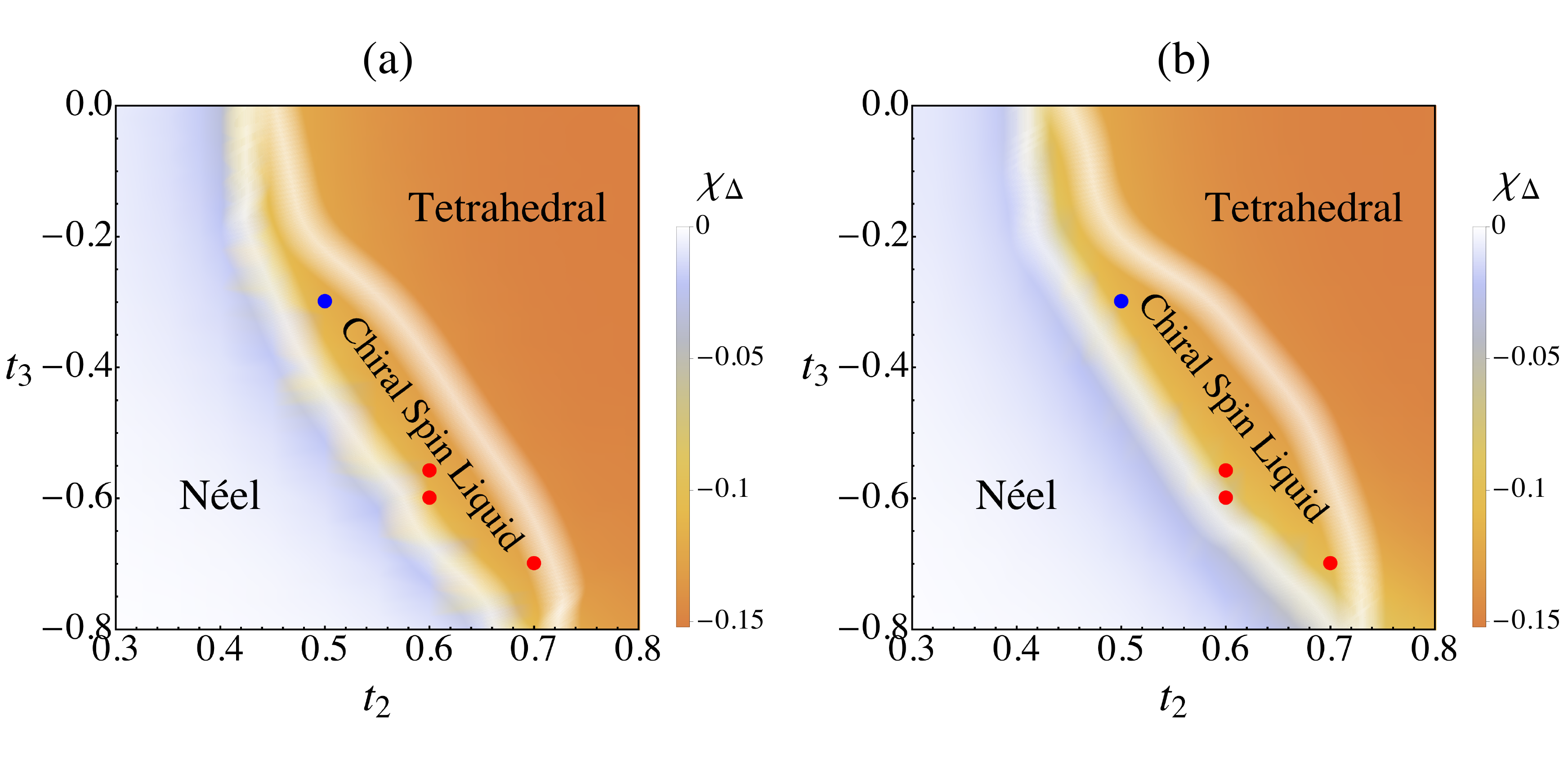}
\caption{ED phase diagram at $\phi=\pi/3$ and $U=10$ for a $N=24$ spin cluster keeping (a) only the additional Heisenberg exchange $J_3$ and (b) all additional terms, 
Heisenberg $J_3$ and chiral $J_\chi$, generated by adding third-neighbor hopping with amplitude $t_3$. DMRG computations have been carried out on the four points marked in 
each phase diagram with all of them showing the expected signatures of a CSL. Energy spectra for the blue points are shown below in 
Fig. \ref{fig:ED_32_Spectra}.}
\label{fig:ED_Compare}
\end{figure}

\begin{figure}[thb]
\begin{center}
\includegraphics[scale=0.4]{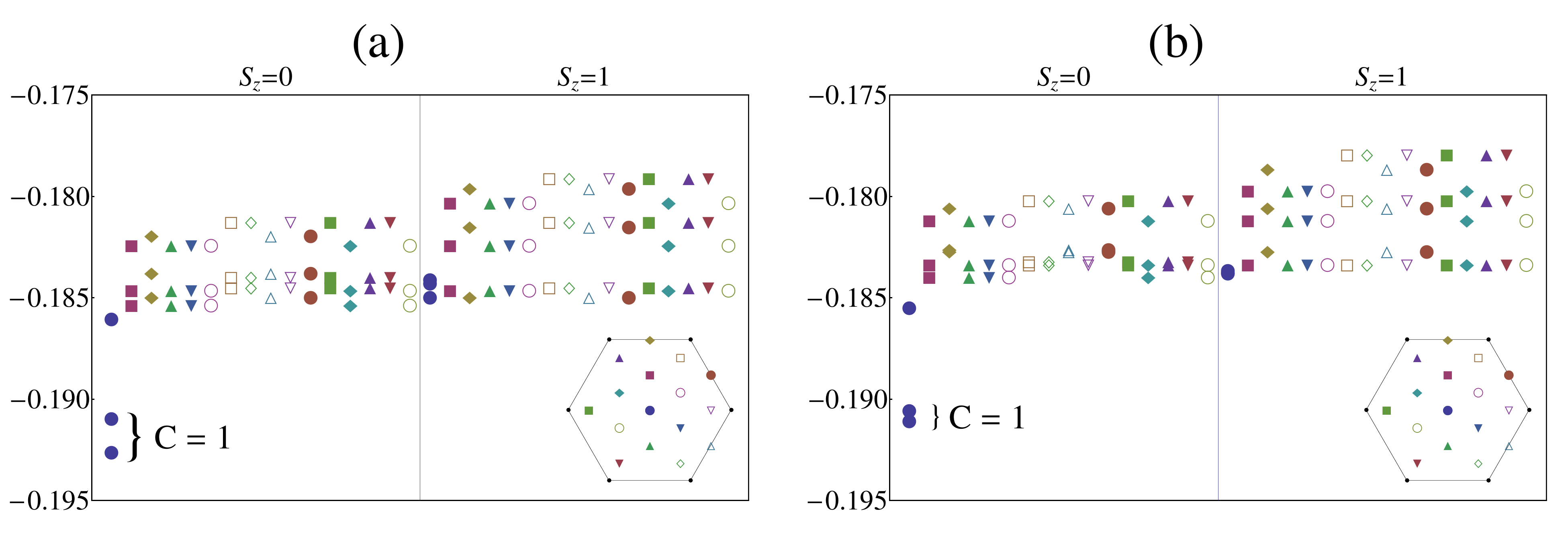}
\caption{Example ED energy spectra for the blue points marked in Fig. \ref{fig:ED_Compare} for a $N=32$ spin cluster with (a) only the additional Heisenberg exchange $J_3$ and (b) all additional terms, 
Heisenberg $J_3$ and chiral $J_\chi$, generated by adding third-neighbor hopping with amplitude $t_3$ . The two lowest lying states in both cases carry total Chern number $C=1$.}
\label{fig:ED_32_Spectra}
\end{center}
\end{figure}

With $t_3=0$, we have found that there is robust magnetically ordered states in the Mott insulating phase of the Haldane-Hubbard model.
With $t_3 \neq 0$, we showed that a CSL phase emerges. In Fig. \ref{fig:ED_Compare} we 
show the phase diagram, at $\phi=\pi/3$, for (a) the case presented in the main text in which only the additional third-neighbor Heisenberg term $J_3 = 4 t^2_3/U$ 
is considered, and (b) the case in which all of the 
additional terms are considered, i.e., the Heisenberg term as well as the additional chiral terms, $J_\chi=24t_1 t_2 t_3/U^2$. In 
Fig. \ref{fig:ED_32_Spectra} we show an example of the energy spectrum for both cases at $(t_2,t_3)=(0.5,-0.3)$. We see that keeping all of the terms results in only very 
small shifts in the phase boundaries, showing that it is really the Heisenberg exchange $J_3$ that is the driving force behind melting the tetrahedral order and getting the CSL phase. 

To reduce the computational complexity of the ED/DMRG computations on the largest system sizes, we have retained only this 
Heisenberg term $J_3$ in the key results presented in the main text. However we have also done DMRG computations (on infinite cylinders with widths up to $L_y=6$) at the four points marked in Fig. \ref{fig:ED_Compare}(b) 
retaining all the extra chiral interactions, and confirmed that the CSL phase is robust.

\section{Field Theory of the Spin Crystallization Transition}

In the main text we constructed a field theory of spin-$1/2$ bosonic spinons 
minimally coupled to an Abelian level $k=2$ Chern-Simons (CS) gauge field to describe a continuous CSL-tetrahedral transition. 
The action is $S \!=\! \int d^2xd\tau ({\cal L}_{\rm CS,\phi} + {\cal L}_{\rm int})$, with
\bea
{\cal L}_{\rm CS,\phi} \! &=& \!\! \frac{1}{2\pi} \epsilon^{\mu\nu\lambda} a_\mu \partial_\nu a_\lambda
\!+\! |(\partial_\mu \!\!-\! i a_\mu) \phi^\pdg_{i\alpha}|^2 \!+\! r |\phi^\pdg_{i\alpha}|^2
\\
{\cal L}_{\rm int} \! &=& \! u_1 (\sum_i \rho_i)^2
\!+\! u_2 \! \sum_{i\neq j}  \rho_i \rho_j
\!+\! u_3 \! \sum_{i\neq j}  {\cal S}_i \cdot {\cal S}_j \!+\! u_4 \!\! \sum_{[ijk\ell]}
\phi^*_{i\alpha} \phi^*_{j\beta}\phi^\pdg_{k\alpha}\phi^\pdg_{\ell\beta}
\!+\! u_5\! \sum_{i\neq j}  \phi^*_{i\alpha} \phi^*_{i\beta}\phi^\pdg_{j\alpha}\phi^\pdg_{j\beta}  \nonumber\\
&+& w_1 (\sum_i \rho_i)^3 + w_2 \sum_{i,j,k} \epsilon^{ijk} {\cal S}_i\cdot {\cal S}_j \times {\cal S}_k + \ldots,
\eea
where Latin indices label the $4$ modes at $\bQ_i$  ($i=0,1,2,3$), $[ijk\ell]$ implies all $4$ modes are different,
there is an implicit sum on Greek indices which label spin or space-time, and
we have defined
$\rho_i \equiv \phi^*_{i\alpha} \phi^\pdg_{i\alpha}$ and 
$\vec {\cal S}_i \equiv \phi^*_{i\alpha} \vec\sigma^\pdg_{\alpha\beta} \phi^\pdg_{i\beta}$.

\begin{figure}[thb]
\includegraphics[scale=0.17]{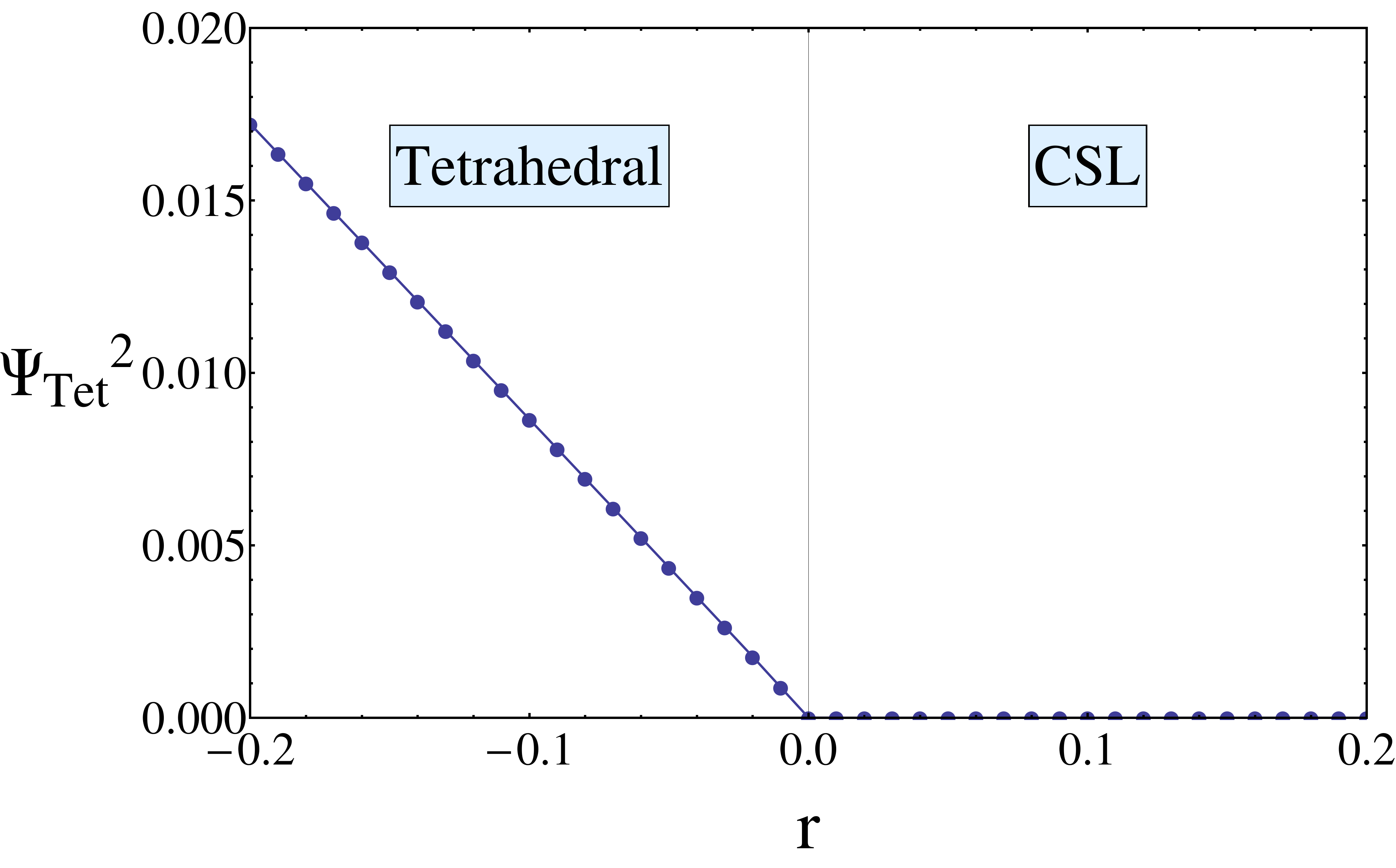}
\caption{Mean field computation of the squared order parameter $ \Psi_{\rm Tet}^2$ for the tetrahedral state within the Chern-Simons-Higgs
field theory, for parameter values in the action mentioned in the text, showing a continuous mean field transition at $r=0$.}
\label{fig:CS_PD}
\end{figure}

At mean field level, we drop all gradient terms.
With dominant $u_1,w_1 \! > \! 0$ and with $u_2<0$, we find $ r  \!>\! 0$ leads to the CSL with $\la \phi_{i\alpha} \ra \! = \! 0$, while
tuning $r \!<\! 0$ leads to a transition into a confining Higgs phase with $\la \phi_{i\alpha} \ra \! \neq \! 0$. 
The tetrahedral state emerges for
subdominant terms $ \! u_4,u_5 \! < \! u_3,w_2$. Fig. \ref{fig:CS_PD} illustrates a concrete example of such a transition, with 
the square of the tetrahedral order parameter,
plotted as a function of $r$ at $u_1=w_1=1$, $u_2=-0.7$, $u_3=w_2=0.1$ and $u_4=u_5=0.01$. It exhibits clear linear 
scaling as expected for the square of a mean-field order parameter. 

\end{document}